# BICYCLING AS A MODE OF TRANSPORT IN DHAKA CITY – STATUS AND PROSPECTS

S. M. Haroon[*], A. K. Bhakta, M. Shahabuddin, N. Rahman & M. R. Mahmud

*Department of Civil and Environmental Engineering, North South University, Dhaka, Bangladesh*
*\*Corresponding Author: saquib.haroon@northsouth.edu*

**ABSTRACT**
This study aims to find out the current status and prospects of using a bicycle as a mode for commuting within Dhaka city. Bicycling is a very sustainable mode of transport but unfortunately is used very less by the commuters of Dhaka. There has been a lot of factors affecting the choice of bicycle to commute. This study was aimed to find out what factors could motivate the commuters of Dhaka to use a bicycle as a mode of transport. For determining the motivators, a survey was administered among the commuters of Dhaka city in which the respondents were asked to answer how certain factors would affect their choice to use a bicycle to commute. A Likert scale was used in the survey and the responses were analysed, from which the top motivators were found. The Motivators were then grouped together using exploratory factor analysis to support possible policymaking. Four factors were extracted using the method. The factors were named Additional Perks, General benefits, Personal Benefits, and Infrastructural benefits

**INTRODUCTION**
The population of Dhaka has been increasing exponentially which has led the city to be tagged as one of the most densely populated cities in the world with a density of 1100 people per square kilometer (Chowdhury et al., 2015). The population of Dhaka is estimated to increase up to 35 million people by 2035 from the current 20 million people (BSS, 2017). The number of motorized cars has increased from 28,764 registered vehicles in 2000 to 1,72,484 in 2011 while on average about 7,000 casualties are reported per year (Ahmed et al., 2014). With increase in population, transportation problems like traffic jams, noise pollution, accidents and other environmental and subjective problems has become a major problem for the city dwellers. For solving these transportation problems bicycle poses to be a sustainable mode of transport (Verma, et al., 2016). Bicycling as a daily travel option also helps in individual as well as environmental health with the cost being nearly negligible (Cavill & Davis, 2007). Bicycling offers an affordable transportation system, less congestion, less pollution and can be used for distances that are long to walk for city dwellers. Despite all these advantages only about one percent of the total dweller in Dhaka city uses bicycle as a mode to commute (Hoque & Alam, 2002).
The deterrents of bicycling include long trip distances of commuters, harsh weather conditions, difficulty to use it in nonutility trips, infrastructure unavailability, extreme traffic conditions, and lack of health and environment consciousness among people (Verma et al. 2016). In the Bangladeshi context, very few researches have been done to understand what acts as motivator or deterrent to bicycling for the people in Dhaka. Bicycle as a medium hasn't been a subject of serious study for the researchers in Dhaka. This study aims to find out the possible motivators that could motivate the commuters of Dhaka to bicycle.

**LITERATURE REVIEW**
Pucher (Pucher, 1997) stated that the main difference in the bicycling mode share between North American cities and the European cities was because of the public policy. Few bike lanes, unmaintained traffic lanes and no respect behaviour for bicyclists were one of the major reasons for not using bicycles

(Pucher, 1997) which is pretty similar to the reasons said by the people who are willing to bicycle in Dhaka. Dill and Voros (Dill & Voros, 2006) identified roads having too much traffic and absence of bicycling lane as two most important barriers for people to bicycle in the Portland region. Quaium (Quaium, 2013) conducted a poll among cyclists of Dhaka and found out that lack of safe parking places, bicycle theft and safety on roads are the three main problems faced by the cyclists in Dhaka city. Dey et al (Dey et al., 2014) conducted a survey among the non-cyclists of Dhaka city and found out six main reasons acting as deterrents for non-cyclists. They were safety, security, parking, long distance travel, social acceptance and discomfort.

Pucher et al (Pucher et al., 2010) reviewed the bicycling policies of different countries and stated that programmatic interventions aim to increase bicycling through promotional activities, media campaigns, educational events, and other means. Bandhan Bandhu Majumdar and Sudeshna Mitra (Majumdar & Mitra, 2015) conducted a study in the Kharagpur city and found out that Physical fitness, Environmental awareness, Travel reliability, Travel flexibility, Psychological safety, Affordability and Desire for pollution free road were the main motivators to bicycling in Kharagpur city of India. Heinen et al (Heinen et al., 2011) conducted a survey among the residents of different cities and conducted a factor analysis concluding that awareness, direct trip-based benefits and safety were the 3 most important motivators to cycling. Verma et al (Verma et al., 2016) conducted a survey among the residents of Bangalore and concluded that stoppage of bicycle during a commuter's childhood was the most important deterrent for commuters to stop using vehicle. They also concluded that cycling lanes and signals at intersections would be strong motivators for people to cycle. Winters et al (Winters, Davidson, Kao, & Teschke, 2011) conducted a survey among 1402 residents of metro Vancouver with 73 possible motivators and deterrents of cycling and grouped them into 15 factors. The most important factors were safety; ease of cycling, weather conditions, route conditions, and interactions with motor vehicles. They also concluded that location and design of bicycle route were the most important factors.

**METHODOLOGY**

An extensive literature review was done first to identify a suitable analysis method to conduct the research. Based on the literatures of (Verma, Rahul, Reddy, & Verma, 2016) and (Mahmud, Gope, & Chowdhury, 2012) it was decided that exploratory factor analysis was to be used for this research. The possible motivators and deterrents were then identified. Various literature were reviewed and possible motivators were identified.

Next a two sectioned questionnaire was designed and administered from among various potential respondents throughout Dhaka city. The first section of the questionnaire consisted of the socio-economic profile of the respondents while the second part was a five point Likert scale questionnaire of 17 possible motivators and deterrents. The questionnaire was administered in two ways – Firstly a pen and paper survey was conducted throughout Dhaka city among various age group of people. Secondly a questionnaire in the format of Google form was sent out electronically to various potential respondents. The responses were then compiled and sorted. Finally 426 responses were deemed accurate. The likert scale was then given values in the following manner - The rating of "Will strongly motivate me" was given a score of +2, the rating of "Will motivate me" was given a score of +1, the rating of "Neutral" was given a score of 0, the rating of "Less likely to motivate" was given a score of -1 and the rating of "Doesn't matter" was given a score of -2. IBM SPSS Statistics 20 was then used for conducting exploratory factor analysis.

**EXPLORATORY FACTOR ANALYSIS**

The exploratory factor analysis was done on a five step exploratory factor analysis protocol developed by Williams et al (Williams, Onsman, & Brown, 2010).

**Step 1: Checking the suitability of data for analysis**

The suitability of data was determined using the Kaiser-Meyer-Olkin (KMO) measure of Sampling Adequacy/Bartlett's Test of Sphericity. KMO values between 0.8 and 1 indicate the sampling is adequate (Cerny & Kaiser, 1977). Our KMO value of 0.833 indicated sampling was adequate.

**Step 2: Selection of factor extraction method**

William et al (2010) mentioned that for behavioural responses where no priory theory exists, Principal Component Analysis or Principal Factor Analysis is generally used. Based on this, the factors were extracted using the method of Principal Factor Analysis.

**Step 3: Selecting Criteria for Factor extraction**
There are many criteria for selecting criteria for factor extraction-: Kaiser's criteria (eigenvalue > 1 rule), the Scree test, the cumulative percent of variance extracted, and parallel analysis (Williams, Onsman, & Brown3, 2010). In this research the Kaiser's criteria that is eigenvalue less than 1 rule was used. Base on this rule 4 factors were extracted

**Step 4: Selection of Rotation Method**
Rotation is used to maximize high item loadings and to minimize low item loadings. This produces a solution which is more interpretable. Oblique rotation are used in research involving human behaviours. This paper uses the Oblique rotation for factor extraction.

**Step 5: Interpretation and labelling of factors**
After extracting the factors, the factors are labelled appropriately for easy interpretation. Variables with higher loading are considered more important than others.

## RESULTS AND ANALYSIS

The Socio-economic characteristics of the respondents are shown below-:

Table 1: Socioeconomic profile of respondents

| | |
|---|---|
| **Gender** | |
| Male | 62.68% |
| Female | 37.32% |
| **Age** | |
| Less than 18 | 2.35% |
| 18-30 | 72.3% |
| 30-45 | 24.18% |
| 45 or more | 1.17% |
| **Bicycle Ownership** | |
| Own a Bicycle | 23.94% |
| Doesn't own a Bicycle | 76.06% |
| **Knows How to Ride a Bicycle** | |
| Yes | 84.98% |
| No | 15.02% |
| **Mode of Transport Currently Used** | |
| Bicycle | 3.52% |
| Others | 96.48% |

Based on the responses of the respondents, the mean was found for each motivators for all the respondents. The factor that "Cycling will protect the environment" was found to be the major motivator with a mean score of +1.63. This was followed by the factor that "Cycling would keep the respondent physically fit" which had a mean score of +1.62. The third most important motivator was that the respondent would cycle if the cycling route had less traffic and air pollution with a mean score of +1.5.

**Results of Exploratory Factor Analysis**
An Exploratory Factor Analysis was done in IBM SPSS Statistics 20 for the recorded response. The Principal Axis factoring method was used for factor extraction while oblimin rotation was used as Rotation technique.
Based on eigenvalues greater than one, four factors were extracted. The factors were named Additional Perks, General benefits, Personal Benefits, Infrastructural benefits. The eigenvalues were 6.504, 1.838, 1.281 and 1.005 respectively. All loading values below 0.3 were not taken into consideration. Additional Perks had the greatest eigenvalue indicating that commuters will be more motivated towards cycling if they are given perks like separate lane, free parking etc. that are not available in other modes of transport. This factor clearly indicates the willingness of commuters to use bicycle. The second factor "General benefits" had an eigenvalue of 1.838. This indicates that people are willing to cycle as it provides environmental benefits as well as other benefits such as less accidents and less traffic prone areas. The third factor extracted was "Personal Benefits" which had an eigenvalue of 1.281. These are the benefits a cyclist would get if he has the facilities to cycle. The last factor extracted was the Infrastructural benefits which are two very important factors for design and should be taken into account while building cycling friendly roads.

Table 2: Exploratory factor analysis results
Extraction Method: Principal Axis Factoring.
Rotation Method: Oblimin with Kaiser Normalization.

| Factors (Eigenvalues) | Variables | Loading Value |
|---|---|---|
| **ADDITIONAL PERKS (6.504)** | | |
| | The cycling lane will protect cyclists from rain | .807 |
| | The cycling route will have very good pavement surface | .767 |
| | There are cloth changing facilities at bicycle parking | .710 |
| | There are separate parking spaces for bicycle | .652 |
| | There is a separate lane for bicycle | .547 |
| | The bicycling route has less intersections | .497 |
| | The parking for bicycle will be free | .393 |
| **GENERAL BENEFITS (1.838)** | | |
| | The cycling route will have less traffic | .682 |
| | Cycling will keep me physically fit | .629 |
| | Cycling will help to protect Environment | .544 |
| | The cycling route is less accident prone | .366 |
| **PERSONAL BENEFITS (1.281)** | | |
| | Cycling will reduce my travel cost | .786 |
| | Cycling will reduce my travel time | .582 |
| | I will be able to travel flexibly using bicycle | .473 |
| | Bicycle is psychologically safer than other modes | .332 |
| **INFRASTRUCTURAL BENEFITS (1.005)** | | |
| | The cycling lane is smooth | .727 |
| | The cycling lane is wide | .658 |

**DISCUSSIONS**

This paper tried to find motivators that could motivate the commuters of Dhaka to bicycle. The paper identified 17 possible motivators and asked respondents to respond how likely he factor could motivate the respondents to bicycling. It should be noted that most of the commuter's (85%) in Dhaka knew how to ride a bicycle and have rode a bicycle but no longer ride a bicycle. This might be because of the weak infrastructural facilities or lack of people around bicycling. One of the most important way to motivate people to bicycling is to motivate people to keep on bicycling even as they grow older.

From the exploratory factor analysis it can be seen that giving perks to bicyclists could motivate people more to bicycle. Options like free parking to bicyclists, separate lanes etc. could motivate commuters to bicycle.

The second most important factor that could motivate people to cycling was General benefits. This included factors like environmental awareness, physical fitness, Traffic & air pollution free roads, and accident free routes. If policies are made to motivate people cycling, the number of commuters using bicycle in Dhaka would increase.

The third most important factor was Personal benefits. This included travel cost, travel time, travel flexibility and psychologically safer roads. If cycling lanes could be made efficient to reduce travel time and make them psychologically safe, a lot of commuters would be motivated to cycling.

The last factor that was extracted was Road conditions. This factor consisted of speed barrier free roads and wide lane. If the cycling lanes are made wide enough and without any interruptions, People would be motivated to bicycling. Policies should be made in such a manner that while constructing roads, road width and interruption free road conditions be enforced.

**CONCLUSIONS**
This paper tried to identify the different motivators that will impact the choice of commuters to use bicycling as mode of transport for commuting in Dhaka city. Four different motivators were identified from the research. The motivators are Additional Perks, General Benefits, Personal Benefits and Infrastructural Benefits. The following conclusions can be made -:

- Although 85% of the respondents learnt to bicycle in their younger days, only 3.6% of these respondents use bicycle as a mode to commute at present. Further research needs to identify the reason for this decline.
- Providing Additional Perks to Bicyclists will be a major factor to motivate non-bicyclists to cycle. These perks include better road pavements, separate parking facilities, free parking and many more. These factors should be considered the most in future urban planning.
- It was identified that general benefits and personal benefits itself will act as motivators for commuters to bicycle. Running well organised campaigns and promoting bicycling itself might act as motivator to commuters.
- A better Infrastructure facility will also motivate the commuters of Dhaka. A separate bicycling lane which is smooth as well as wide will highly motivate the commuters of Dhaka to bicycle. Urban planners, engineers and policy makers need to change their designs and policies accordingly to promote bicycling.

Dhaka has a very less mode share for bicycles unlike countries like Netherlands, Germany and Canada. The transportation system of Dhaka city is termed as worst in few cases. Integrating bicycling as a mode of transport within the current transportation system might be of great impact. Very few studies were found regarding bicycling as a mode of transport in Dhaka city. Further research on bicycling is needed to unearth potential of bicycling in Dhaka city.

**ACKNOWLEDGMENTS**
We gratefully thank all the 426 respondents of this survey who have provided us with valuable insights for this research. We would also like to thank all the anonymous referees who provided us with valuable suggestion for this research.